\documentclass[11pt]{article} 

\usepackage[utf8]{inputenc} 

\usepackage{geometry} 
\usepackage{graphicx} 

\usepackage{booktabs} 
\usepackage{array} 
\usepackage{paralist} 
\usepackage{verbatim} 
\usepackage{subfig} 

\usepackage{pifont}
\usepackage{amssymb}

\usepackage{fancyhdr} 
\usepackage{sectsty}
\allsectionsfont{\sffamily\mdseries\upshape} 

\usepackage[nottoc,notlof,notlot]{tocbibind} 
\usepackage[titles,subfigure]{tocloft} 

\title{Hot RAD: \\ A Tool for Analysis of Next-Gen RAD Tag Data}
\author{Lauren A. Assour, Nicholas LaRosa, and Scott J. Emrich}

\begin{document}
\maketitle

\begin{abstract}
Restriction site Associated DNA (RAD) tagging (also known as RAD-seq, etc.) is an emerging method for analyzing an organism’s genome without completely sequencing it. This can be applied to a non-model organism without a reference genome, though this creates the problem of how to begin data analysis on unmapped and unannotated reads. Our program, Hot RAD, presents a straightforward and easy-to-use method to take raw Illumina data that has been RAD tagged and produce consensus contigs or sequence stacks using a distributed framework, creating a basis on which to begin analyzing an organism’s DNA. The GUI (graphical user interface) element of our tool makes it easy for those not familiar with the command line to take raw sequence files and produce usable data in a timely manner.
\end{abstract}

\section{Introduction}
Exploring the genetic makeup of diverse non-model organisms has always posed something of an interesting problem. With no reference genome, scientists are required to assemble complex data \textit{de novo} in a fashion that is time-consuming and requires countless iterations of differing parameters to get the best results. As sequencing technology improves researchers are also left with more reads to manage, creating multiple bottlenecks in the analysis process.  

One newer genome reduction technique to streamline genetic information sampling is called {\textbf R}estriction site {\textbf A}ssociated {\textbf D}NA tagging, or RAD tagging (sometimes called RAD-seq, among other names)\cite{radtag}. As the name suggests, this approach uses enzymes to cut along an organism's genome at specific restriction sites, after which tags (barcodes) are attached to the enzyme-cut end of the sequence. This allows pooling of individuals for sequencing, and also creates data that has predictable characteristics. The most important (and problematic) of these characteristics is that all sequences are cut by the same enzyme, and thus are all from the same locations in the genome. Unlike traditional sequencing, reads should stack on top of one another as opposed to the usual staggered shotgun sampling expected by current assembly programs.

Because of this difference traditional assemblers, while feasible solutions, are not the optimal tool for RAD tag analysis. We have created a pipeline which, unlike traditional assemblers, generates either clusters of sequences or a consensus sequence for each restriction site location to which other RAD samples can best be aligned to. This has been packaged into a GUI (graphical user interface) for ease of use by non-expert users, though the entirety of the pipeline can also be run through command line execution of the underlying code. To deal with large data sets, our program also can be run on multiple distributed systems/clouds using Makeflow and Weaver\cite{makeflow}\cite{weaver}. More importantly our tool, called Hot RAD, utilizes more sequences in its alignment, sorting them into fewer, deeper loci with less memory usage as compared to other RAD tag analysis methods.

\begin{figure*}
\caption{Hot RAD Pipeline}
\label{fig:hrpipeline}
\centering
\includegraphics[width=16cm]{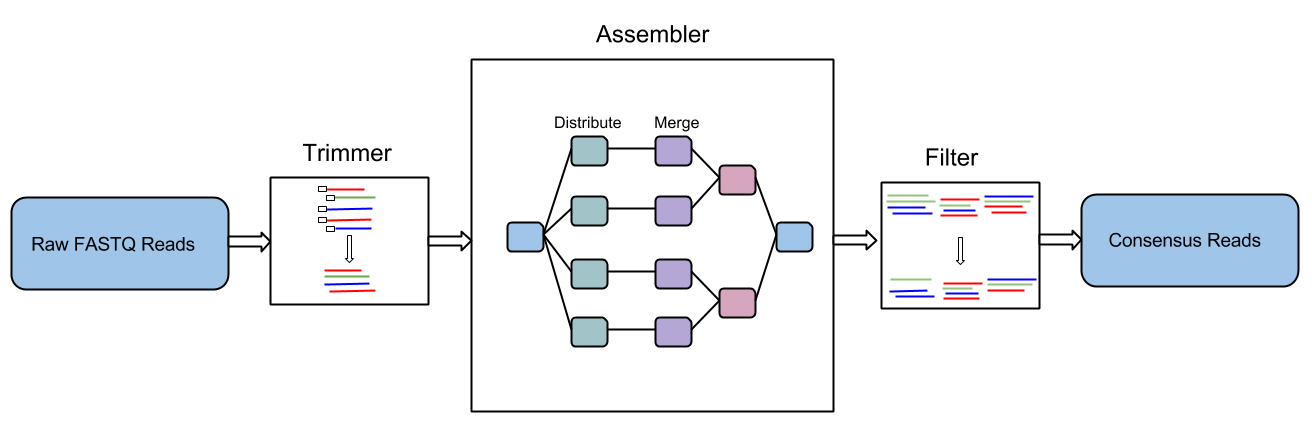} 
\end{figure*}

\section{Methods}
The GUI is constructed in Java while the functionality of the program is encompassed in Python scripts, Java code, and Makeflow and Weaver scripts. The GUI allows users to choose what functions to carry out on the data and what values to set parameters at. It then lets the user choose between running directly from the GUI or constructing a shell script to run all of the chosen steps on whichever system/cloud they have access to. Any step in Hot RAD can be left out or substituted with another program, allowing users to only utilize what they need. For example, a different alignment tool (such as an alternative assembler) can be utilized in place of the provided assembler described below.

\subsection{Trimming}
The sequence trimmer is implemented in Python and preps the raw FASTQ data to be processed. It allows users to input the sequenced end of the cut site and a barcode list so that sequences can be trimmed (removing barcodes and the repetitive cut site end) and labeled, preparing them for downstream analysis. This portion of the program includes barcode correction and removes all reads that have no valid barcode or no valid cut site. Barcodes are not required to all be of the same length. Minor quality control is available through the option to remove sequences with greater than a given number of Ns. Paired-end reads can be provided to this step; the trimmer removes all secondary reads whose primary reads are dropped for too many Ns, no cut site, or an unidentifiable barcode. An option to remove reverse reads that are read-through of the forward read is also available.

\subsection{Alignment}
The first step in the assembly process involves grouping sequences based on the percent identity value provided by the user. The percent identity between sequences is calculated using the Smith-Waterman algorithm, a dynamic programming approach that allows us to estimate the maximum number of matched nucleotides in an alignment between two sequences\cite{smithwaterman}; to decrease run time the banded version of this algorithm has been implemented\cite{banded}. An option for customized Smith-Waterman scoring schemes is included for added control. During the alignment, those sequences that do not match with any other groups are put into their own singleton group until every sequence is either grouped with similar reads or is in a singleton group. This is distributed using a Weaver-based MakeFlow, and can be run locally or over a distributed system like condor with WorkQueue\cite{condor}\cite{workqueue}. 

The second step involves finding the most representative (consensus) sequence using a modified center star tree alignment process\cite{gusfield}. This process again utilizes the Smith-Waterman algorithm to determine which haplotype sequence within a group leads to the lowest alignment score (least edit distance) when aligned with every other sequence in that respective group. This consensus haplotype is the equivalent of a contig from an assembler, and is output in FASTA format followed by all other sequences in its group.

The FASTA file of consensus alleles is then re-aligned using the same grouping process seen in the first step to remove any duplicate groups or ungrouped singletons that may have locus representatives in other files, an artifact of the distributed nature of the first step. This is done in a tiered merging fashion, as seen in the Assembler step of Figure \ref{fig:hrpipeline}. Starting from the original output of each distributed step, files are merged using Smith-Waterman, grouping similar consensus sequences and their groups together. In the final merge, users can specify removal of groups with low sequence coverage to speed the process, such as removing all singletons. Once all consensus sequences are grouped, all consensus sequences and their groups are output in FASTA format. 

\begin{table}[!t]
\centering
\caption{RUN TIMES OF HOT RAD WITH DIFFERING FILTERS}
\label{tab:hrfilts}
\begin{tabular}{llll}\toprule\hline
\textbf{Filters} & \textbf{Run Time (Min)} & \textbf{Reads Used} & \textbf{Clusters}\\\midrule
None & 773.1 & 21,940 & 4,437\\
BP Count & 169.2 & 22,010 & 4,480\\
BP + Fast Align & 16.9 & 18,524 & 3,108\\
BP + FA + Banded & 6.6 & 18,257 & 3,106 \\
\bottomrule
\end{tabular}
\\[6pt]NOTE: BP Count refers to counting the occurrence of each base. Fast Align refers to checking for similarity in the initial bases of the sequences. Run using 100 jobs and workers on condor with Work Queue.
\end{table}

We note that commonly used exact-match speedups can be used in some cases, but we are concerned with natural populations that could potentially be only 90\% similar over 110bp (implying exact matches as small as 10bp). To speed up run time in a memory-efficient manner, a simple filtration step derived from Gusfield has been added\cite{gusfield}. Given that we expect the sequences from a locus to be highly similar, we can expect the nucleotide frequencies to be very similar as well, even if these sequences possess one or more insertions or deletions. Therefore, we compute a basic count of the occurrence of each nucleotide and if the sum of the absolute values of the differences is greater than a user provided value, we do not attempt to align these sequences. This is in effect a linear time estimate of edit distance.

Another filtration step is also available to reduce Hot RAD's run time. Because of the nature of RAD data, users can assume that their sequences will stack with a relatively small offset at the beginning of sequences. Insertions and deletions immediately following cut sites should be rare enough that we can expect the initial portions of sequences to match up relatively close to one another. Because of this, we can add an extra layer of filtration to prevent attempting to align sequences that likely won't pass the percent identity cutoff because of disparate initial bases. Hot RAD aligns the first X number of bases to a maximum offset Y, relying on the knowledge that reads that belong to the same locus should have a relatively good alignment in the initial bases. The run time advantages to these described filters can be seen in Table \ref{tab:hrfilts}, which compares these filters when used with standard Smith-Waterman alignment and finally with banded alignment. 

\subsection{Filtration}
This step allows users to filter out consensus sequences or contigs (if an assembler was used) that do not meet certain criteria. Because of the nature of RAD tag data, constructed contigs or consensus sequences should not be much longer or shorter than the longest and shortest reads, and should have a moderate amount of coverage overall. Contigs can be removed as erroneous using these settings. If the provided assembler is used, consensus reads will be of appropriate lengths, but the option remains to filter further.

Ultimately, the output is a set of consensus sequences with the option of including all of the sequences that appear in their respective groups. Though the content of the output varies somewhat depending on options chosen and alignment tool used, Hot RAD is highly flexible regardless.

\section{Results}
\subsection{Data Sets}
Singled-ended Illumina RAD reads from \textit{Rhagoletis pomonella} flies provided by the Feder lab at Notre Dame are the primary set used in this paper.  These were run on multiple platforms (Illumina HiSeq and Illumina GA II) at NCGR (Arizona) \cite{exprhag}. Our results show correct sequence grouping and high sequence usage when compared to other approaches.

\subsection{Hot RAD}
We ran Hot RAD with differing numbers of workers, numbers of sequences, and sources of data to profile the performance. Given the relatively few number of loci sampled relative to overall dataset size \cite{exprhag}, most experiments were done on subsets on the order of tens to hundreds of thousand sequences.  Overall Hot RAD performed well, generating results that were used as the reference loci used in  \cite{exprhag}.

Table \ref{tab:hrreadusage} shows differing sequence usage (e.g. how many sequences were placed into a putative RAD locus) and consensus sequences between Stacks, Hot RAD, and SeqMan NGen, a proprietary assembler discussed further below. Stacks utilized default parameters (minimum depth of 2 sequences per stack). Hot RAD and SeqMan used minimum match percentage 90\%, match score 10, mismatch -15, gap -40 and a minimum depth of 2 (minimum depth of 3 for Hot RAD*). Hot RAD's filtration steps outlined above used a maximum count difference of 5 bases in a 60 base window, and a match of at least 5 bases in a search area of 8 bases at the beginning of the sequence, with a maximum offset of 3 bases. Sequences in groups of one (not clustered with any other reads) were removed before the final merge step. SeqMan's numbers were taken before the filtration step to remove contigs that do not fit length requirements.

Table \ref{tab:hrfilts} gives a basic indication of Hot RAD's performance and run time using different filtering options to decrease run time. We show that Hot RAD performs well with highly polymorphic data, and can manage sequences of varying lengths, quality, and sequences that have gaps or insertions in relation to other sequences at the same locus. By using relatively more computing resources it can utilize more sequences consistently when compared to other approaches like Stacks and SeqMan with similar settings.

\begin{table}
\renewcommand{\arraystretch}{1.3}
\caption{USAGE OF 200K RAD READS BY PROGRAM}
\label{tab:hrreadusage}
\centering
\begin{tabular}{llll}\toprule\hline
\textbf{Program} & \textbf{Reads Used} & \textbf{\% Used} & \textbf{Clusters}\\\midrule
Hot RAD & 100,250 & 50.1\% & 17,854 \\
Hot RAD* & 78,503 & 49.3\% & 7,043 \\
Stacks & 75,737 & 37.9\% & 14,124 \\
SeqMan & 100,711 & 50.4\% & 30,389 \\
\bottomrule
\end{tabular}
\\[3pt]
NOTE: Banded Hot RAD was utilized for this table. *Required a minimum depth of 3 sequences per locus. 
\end{table}

\begin{table}
\caption{RUN TIME OF 50K RAD SEQUENCES BY PROGRAM}
\label{tab:hrruntime}
\centering
\begin{tabular}{lll}\toprule\hline
\textbf{Program} & \textbf{Run Time (Min)} & \textbf{Cores (Local)} \\\midrule
Hot RAD & 6.6 & 8 \\
Stacks & 1.1 & 1 \\
SeqMan & .267 & 1 \\
\bottomrule
\end{tabular}
\end{table}

\subsection{Previous Approaches}
As mentioned previously, different methods have been used to analyze RAD tag data without a reference genome, ranging from using pre-existing assemblers to writing new tools. Here we outline these alternative methods while highlighting their differences with Hot RAD. Table \ref{tab:radapproaches} gives a brief overview of capabilities of different RAD tag analysis approaches.

One application for RAD analysis is called Rainbow\cite{rainbow}. This application only takes paired-end RAD data to be processed. It is single-core and does not utilize `overhanging bases,' so the shortest read dictates the longest stretch of base pairs that are aligned. It utilizes the second sequence in paired-end sequences to better inform its alignment of reads. It uses a large amount of memory during its execution.

Outlined in the Rainbow paper is another approach called RApiD\cite{rapid}. Though their paper provides no run time information, it is significantly slower than Rainbow. Unlike Rainbow, it is simply a suite of scripts that utilize outside programs to do meaningful work on the data. It requires use of the program Vmatch, which provides free licenses for non-commercial research institutions\cite{vmatch}. It also only takes paired-end reads.

One of the first programs for RAD data, Stacks, was written by the lab that developed the RAD tag method\cite{stacks}. It is distributed across cores and takes both single-end and paired-end reads. Unlike the other approaches it also provides scripts to demultiplex reads. Similar to Rainbow it can only work with a single read length, and thus requires users to trim sequences to the same lengths or else these different length reads will not be used. Stacks are generated based on maximum number of mismatches, and thus cannot handle insertions or deletions in a sequence in a given locus. Stacks has a large memory footprint from loading all sequences into memory, and runs rather slowly on large data sets. 

\begin{table*}
\centering
\caption{APPROACHES TO ANALYZING RAD DATA}
\label{tab:radapproaches}
\begin{tabular}{llllllll}
\toprule\hline 
 & \textbf{RAD}  & \textbf{Single} & \textbf{Paired} & \textbf{Threads/} &\textbf{Different}  & \textbf{Quality} \\
\textbf{Program} & \textbf{Specific} & \textbf{End} & \textbf{End} & \textbf{Distributed} & \textbf{Lengths} & \textbf{Scores} \\
\midrule
Stacks & \checkmark & \checkmark & \checkmark & \checkmark (Threads) & \ding{55}  & \ding{55}  \\
Rainbow & \checkmark & \ding{55} & \checkmark & \ding{55}  & \ding{55}  & \ding{55}  \\
RApiD & \checkmark & \ding{55}  & \checkmark & \ding{55}  & \checkmark & \ding{55}  \\
RADTools & \checkmark & \checkmark & \checkmark & \checkmark (Threads) & \ding{55}  & \ding{55}  \\
SeqMan & \ding{55}  & \checkmark & \checkmark & \ding{55}  & \checkmark & \checkmark \\
\textbf{Hot RAD} & \checkmark & \checkmark & \checkmark* & \checkmark & \checkmark & \ding{55}  \\
\bottomrule
\end{tabular}
\\[3pt]
NOTE: * Hot RAD can process paired-end data but does not use the second reads to assist in read placement.
\end{table*}

RADTools is an approach out of the University of Edinburgh\cite{radtools}. The developers recommend using Stacks in place of their program when possible. It can use single-end or paired-end reads. Like Stacks, RADTools generates groups based on maximum number of mismatches and can run multithreaded. RADTools provides a demultiplexing tool like Stacks and Hot RAD, but it is very rigid. It is unable to handle variable length barcodes, or at the very least only looks at a specific start index for the cut site. This results in many reads being discarded because the cut site `does not match,' which can leave researchers with a very small subset of their data. To use a demultiplexing tool other than the one provided requires some modification given that RADTools do not support basic FASTQ, but a slightly modified version of the format.

Hot RAD has the option of using a different assembler or alignment tool instead of the one included. Originally this is how the pipeline was run, relying on a proprietary assembler called SeqMan NGen\cite{seqman}. One drawback of using an assembler is that it works with RAD data like it does normal sequences and tries to stagger and join them together into longer reads called contigs. SeqMan is not distributed and takes both single-end and paired-end reads, but as suggested requires some post-processing (included in Hot RAD) to remove contigs that are not useful because they fall out of the range of expected sequence stacks. It has good data usage but requires a paid license, although a 30-day free trial can be obtained by academics. SeqMan NGen will only places reads in a single contig, but it also sometimes only aligns a portion of a read instead of its entirety, leading to erroneous cluster assignments. 

\section{Future Work}

Although to our knowlegde Hot RAD is unique in its ability to process highly heterogeneous data (oak, field caught flies, and mosquitoes), it would be useful to process paired-end data to improve data usage and clustering where possible.  This framework could benefit from additional improvements.  For example, Hot RAD and none of the approaches listed in Table \ref{tab:radapproaches} use base quality information to inform their sequence alignments, which could better decipher variants during or after clustering.  Moreover, additional or new pre-processing with filters could provide a ``fast'' alignment option for Hot RAD to accompany the slower more accurate settings of the current implementation.  Finally, the preparatory work for \textbf{G}enotyping \textbf{B}y \textbf{S}equencing (GBS)\cite{gbs} is slighly different than traditional RAD because it requires an extra step of trimming trailing adapter sequence base pairs. The modular nature of our implementation will allow developing a GBS-specific version for analyzing these data by simply replacing the current preprocessing steps.

\section{Conclusions}
We have constructed a tool that requires no external pre-processing of the data and that can be further analyzed and utilized by researchers, even if the are unfamilar with command line tools. Its optional distributed nature allows for faster execution, and it can be used on machines without a large memory requirement. Hot RAD consistently uses more sequences than other approaches but puts them into fewer, deeper stacks. It is comparable with other RAD tag analysis software with the added benefit of flexibility, straightforward output, a GUI, and a high level of customization.

\bibliographystyle{plain}
\bibliography{new_hr.bib} 

\begin{thebibliography}{10}

\bibitem{vmatch}
Mohamed~I. Abouelhoda, Stefan Kurtz, and Enno Ohlebusch.
\newblock Replacing suffix trees with enhanced suffix arrays.
\newblock {\em J. of Discrete Algorithms}, 2(1):53--86, March 2004.

\bibitem{makeflow}
M.~Albrecht et~al.
\newblock Makeflow: a portable abstraction for data intensive computing on
  clusters, clouds, and grids.
\newblock In {\em 1st ACM SIGMOD Workshop on Scalable Workflow Execution
  Engines and Technologies, ser}. ACM, 2012.

\bibitem{radtag}
N.~Baird et~al.
\newblock Rapid snp discovery and genetic mapping using sequenced rad markers.
\newblock {\em PLoS ONE}, 3(10):e3376+, 2008.

\bibitem{radtools}
Simon~W. Baxter, John~W. Davey, J.~Spencer Johnston, Anthony~M. Shelton,
  David~G. Heckel, Chris~D. Jiggins, and Mark~L. Blaxter.
\newblock Linkage mapping and comparative genomics using next-generation {RAD}
  sequencing of a non-model organism.
\newblock {\em PloS one}, 6(4):e19315+, April 2011.

\bibitem{weaver}
P.~Bui et~al.
\newblock Weaver: Integrating distributed computing abstractions into
  scientific workflows using python.
\newblock In {\em CLADE}, 2010.

\bibitem{workqueue}
P.~Bui et~al.
\newblock Work queue + python: A framework for scalable scientific ensemble
  applications.
\newblock In {\em ACM/IEEE International Conference for High Performance
  Computing, Networking, Storage, and Analysis (Supercomputing)}, Workshop on
  Python for High Performance and Scientific Computing (PyHPC), 2011.

\bibitem{stacks}
J.~Catchen et~al.
\newblock Stacks: Building and genotyping loci de novo from short-read
  sequences.
\newblock {\em G3: Genes, Genomes, Genetics}, 1(3):171--182, 2011.

\bibitem{banded}
K.-M. Chao et~al.
\newblock Aligning two sequences within a specified diagonal band.
\newblock {\em Computer applications in the biosciences : CABIOS},
  8(5):481--487, 1992.

\bibitem{rainbow}
Z.~Chong et~al.
\newblock Rainbow: an integrated tool for efficient clustering and assembling
  rad-seq reads.
\newblock {\em Bioinformatics}, 28(21):2732--2737, 2012.

\bibitem{seqman}
DNAStar.
\newblock Seqman ngen.

\bibitem{exprhag}
Scott~P. Egan, Gregory~J. Ragland, Lauren Assour, Thomas H.~Q. Powell, Glen~R.
  Hood, Scott Emrich, Patrik Nosil, and Jeffrey~L. Feder.
\newblock Experimental evidence of genome-wide impact of ecological selection
  during early stages of speciation-with-gene-flow.
\newblock {\em Ecol Lett}, page n/a, June 2015.

\bibitem{gbs}
Robert~J. Elshire, Jeffrey~C. Glaubitz, Qi~Sun, Jesse~A. Poland, Ken Kawamoto,
  Edward~S. Buckler, and Sharon~E. Mitchell.
\newblock A robust, simple genotyping-by-sequencing ({GBS}) approach for high
  diversity species.
\newblock {\em PloS one}, 6(5):e19379+, May 2011.

\bibitem{gusfield}
Dan Gusfield.
\newblock {\em Algorithms on Strings, Trees and Sequences: Computer Science and
  Computational Biology}.
\newblock Cambridge University Press, 1 edition, May 1997.

\bibitem{condor}
M.~Litzkow et~al.
\newblock Condor-a hunter of idle workstations.
\newblock In {\em 8th International Conference on Distributed Computing
  Systems}, pages 104--111. IEEE, 1988.

\bibitem{smithwaterman}
T.~Smith and M.~Waterman.
\newblock Identification of common molecular subsequences.
\newblock {\em Journal of Molecular Biology}, 147(1):195--197, 1981.

\bibitem{rapid}
Eva-Maria Willing, Margarete Hoffmann, Juliane~D. Klein, Detlef Weigel, and
  Christine Dreyer.
\newblock Paired-end {RAD}-seq for de novo assembly and marker design without
  available reference.
\newblock {\em Bioinformatics}, 27(16):2187--2193, August 2011.

\end{thebibliography}

\end{document}